\documentclass[final,1p,times]{elsarticle} 
\usepackage{graphicx}
\usepackage{amssymb} 
\usepackage{amsthm} 
\usepackage{lineno}

\journal{Nuclear Physics A} 

\begin{document}

\begin{frontmatter} 

% Your Title - please insert
\title{Cold Nuclear Matter Effects in d+Au Collisions at PHENIX}

%% Single author (and collaboration) - please insert
\author{Baldo Sahlmueller (for the PHENIX\fnref{col1} Collaboration)}
\fntext[col1] {A list of members of the PHENIX Collaboration and acknowledgements can be found at the end of this issue.}
\address{Stony Brook University, University of Frankfurt}

%% Multiple authors
%\author[auth2]{Marcus Junius Brutus}
%\address[auth1]{Somewhere, Rome}
%\address[auth2]{Somewhere else, Rome}

\begin{abstract} 
  To interpret the measurements in heavy-ion collisions at the Relativistic Heavy-Ion Collider (RHIC) it is crucial to understand the initial state of the colliding gold (Au) nuclei. The parton distribution in Au nuclei is modified compared to protons, and their isospin composition is different due to the presence of neutrons. d+Au collisions at RHIC at the same collision energies are an important tool to study initial state modifications. PHENIX has measured $\pi^0$, $\eta$, and reconstructed jets at high transverse momentum. These data are compared to predictions from nuclear parton distribution functions. Furthermore, single electrons from heavy-flavor decays have been measured by PHENIX.
\end{abstract} 

\end{frontmatter} % do not change

%% linenumbers are useful for reviewing process
%\linenumbers

\section{Introduction}

RHIC experiments have established the formation of a quark-gluon plasma (QGP) in heavy-ion collisions at $\sqrt{s_{NN}} = 200$\,GeV~\cite{PHENIXwhitepaper}. Among the crucial probes showing the presence of the QGP is jet quenching, measured as the suppression of high $p_T$ hadrons such as $\pi^0$ and $\eta$. Early measurements in $d$+Au collisions showed the absence of jet quenching in a system where no hot and dense matter is created. Later measurements in Au+Au collisions revealed that single electrons from heavy flavor decays are suppressed as well at high $p_T$. The much larger 2008 $d$+Au data now allow further study of the initial state of heavy-ion collisions, especially concerning the centrality dependence of such effects. The data also allow the reconstruction of jets in PHENIX and the measurement of $\pi^0$ and $\eta$ as well as of heavy-flavor electrons up to significantly higher $p_T$.
Initial state effects in heavy-ion collisions include the so-called Cronin enhancement~\cite{Cronin}, a broadening of the transverse momentum distribution of particles compared to the $p$+$p$ baseline, modifications of nuclear parton distribution functions (PDFs), or the possible formation of a color glass condensate.
Recently, centrality dependent nuclear PDFs, EPS09s and EKS98s, were published~\cite{eps09s}. They are an extension of the prior EPS09 and EKS98 nuclear PDFs with an impact parameter dependence included (based on the $A$ dependence of those PDFs).  They predict a small dependence of the nuclear modification factor in $d$+Au and $p$+Au on the collision centrality.

\section{Data Set and Centrality Selection}

The results presented in these proceedings are from the analysis of the 2008 $d$+Au data taken by PHENIX central spectrometer. PHENIX sampled 80\,nb$^{-1}$ during that run, a factor of 30 increase over the 2003 $d$+Au data set.

The centrality determination is a common, crucial component to all analyses presented here. The collision centrality was determined using the charged multiplicity in forward rapidity $3.1 < \eta < 3.9$, as measured with the PHENIX beam-beam counter. Here, forward (positive) rapidity is defined as the Au-going side. The underlying average number of binary nucleon-nucleon collisions ($\langle N_{coll}\rangle$) was estimated with a Glauber model Monte Carlo simulation.

Due to the autocorrelation between having a high $p_T$ particle in the central PHENIX spectrometer and the charge multiplicity in the BBC, an additional correction factor is calculated. This is done with the same Glauber Monte Carlo simulation used for the calculation of $N_{coll}$.

\begin{figure}[htbp]
\begin{minipage}[hbt]{77mm}
\begin{center}
\includegraphics[width=\textwidth]{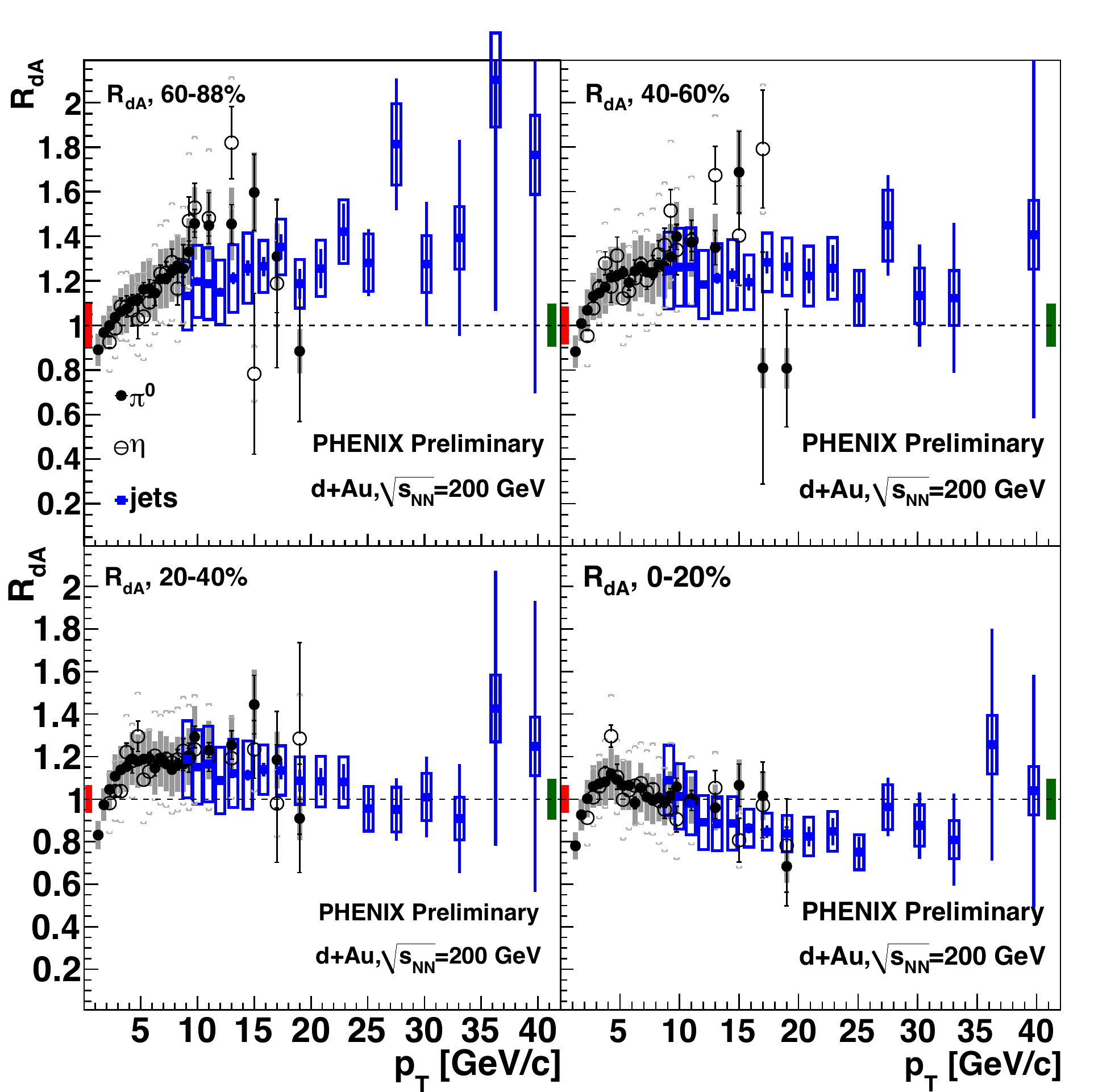}
\end{center}
\caption{Centrality dependence of the nuclear modification factor, $R_{dA}$, for $\pi^0$ (dots), reconstructed jets (squares), and $\eta$ (circles).The boxes around the points show $p_T$ correlated systematic uncertainties, the boxes on the left depict the uncertainty of $N_{coll}$, the ones on the right show the $p$+$p$ normalization uncertainty.}
\label{fig:rda}
\end{minipage}
\hspace{\fill}
\begin{minipage}[hbt]{58mm}
\begin{center}
\includegraphics[width=\textwidth]{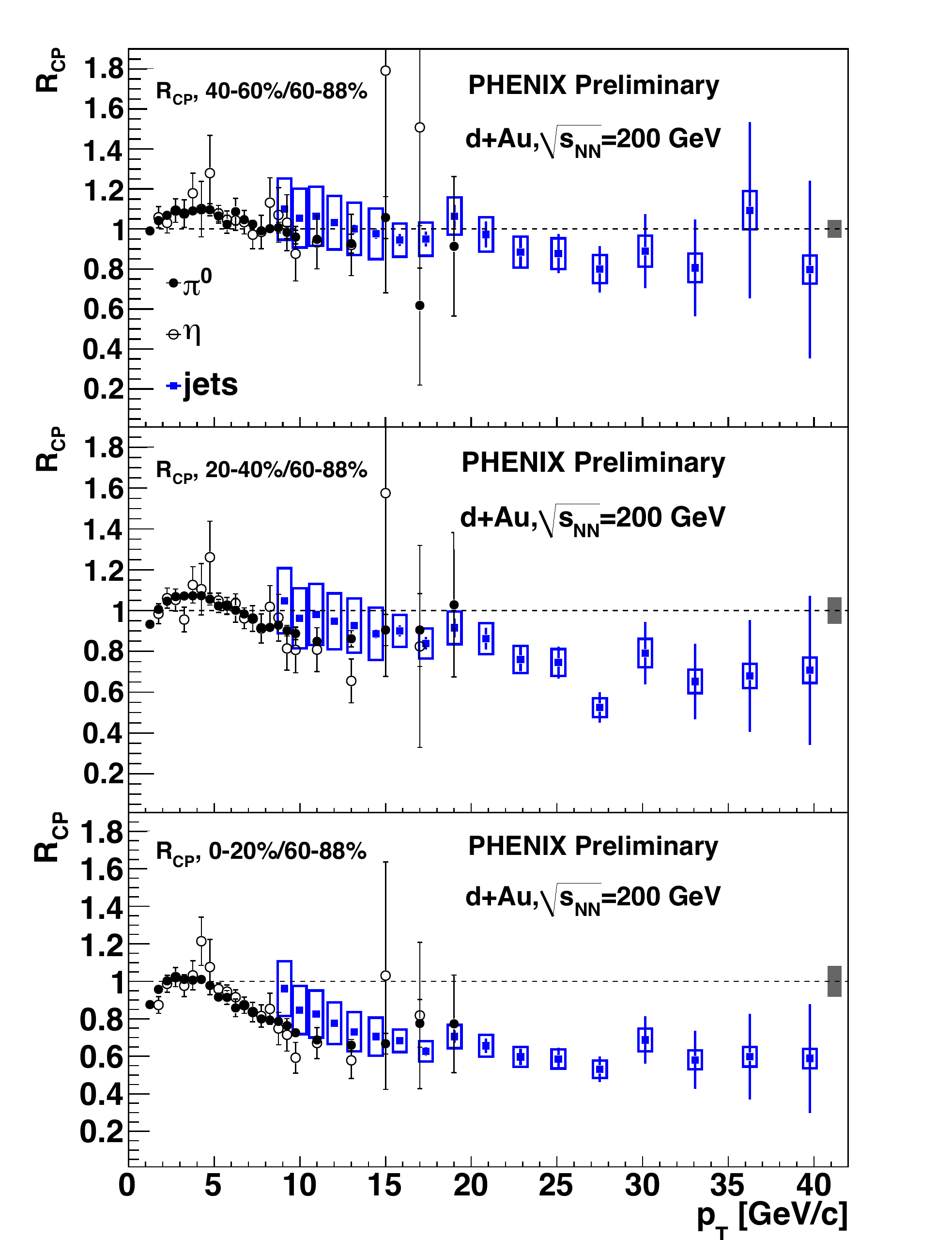}
\end{center}
\caption{Ratio of the yield in more central collisions to most peripheral collisions, $R_{CP}$, for $\pi^0$ (dots), reconstructed jets (squares), and $\eta$ (circles). The boxes show the same systematic uncertainties as in Figure~\ref{fig:rda}.}
\label{fig:rcp}
\end{minipage}
\end{figure}

\section{High $p_T$ $\pi^0$, $\eta$, and Jets}

The analysis method for $\pi^0$ and $\eta$ is based on the earlier PHENIX approach used for the 2003 data~\cite{ppg044}. Since the $p_T$ reach of this analysis is significantly higher, the merging of the electromagnetic showers from the two $\pi^0$ decay photons has to be taken into account. Such showers are removed by a shower shape cut. The probability for this loss of $\pi^0$ in the data sample was estimated with a fast Monte Carlo program that takes into account the shape of the electromagnetic showers of the decay photons.
The jet analysis uses the Gaussian filter algorithm~\cite{gaussian} that was developed in the analyses of $p$+$p$ and Cu+Cu data in PHENIX. The algorithm is a seedless cone-like algorithm with Gaussian weighting on the angle with respect to the jet axis. It is an infrared- and collinear-safe algorithm that was developed for heavy-ion collisions. It focuses on the energetic core of the jet and optimizes the signal to background ratio. The small underlying event present in $d$+Au collisions was evaluated with embedding studies. For the data shown here, the fake rate was determined to be smaller than 5\%.

PHENIX has measured the spectra of $\pi^0$ ($\eta$) for $p_T$ up to 20 (22) GeV/$c$. Since the decay photons of the $\eta$ have a larger separation due to the larger mass of the parent particle, the $\eta$ could be measured up to higher $p_T$ than the $\pi^0$, despite its smaller production cross section and smaller two-photon branching ratio. The jet spectra were measured for 9 GeV/$c$ $<$ $p_T$ $<$ 45 GeV/$c$. The nuclear modification factor $R_{dA}$, is then calculated for each of the three final states, X, as
$$R_{dA} = \frac{dN^{X}_{d+Au}/dp_Tdy}{N_{coll}\cdot dN^{X}_{p+p}/dp_Tdy}\mbox{ .}$$

$R_{dA}$ is shown in Figure~\ref{fig:rda} for four centrality selections, for $\pi^0$, jets, and $\eta$. The systematic uncertainties shown are (in the $\pi^0$ and $\eta$ case) mostly due to using a $p$+$p$ baseline from a different RHIC run with an independent energy calibration of the calorimeter. The $p_T$-correlated systematic uncertainties are mostly correlated between $\pi^0$ and $\eta$ while they are uncorrelated  between the mesons and the jet measurement. In central events, high $p_T$ jets appear slightly suppressed, with $R_{dA}\approx 0.85$, and the $\pi^0$ and $\eta$ show consistent behavior. In peripheral events, however, all three probes are enhanced compared to $p$+$p$ collisions. This behavior is consistent with the observation in an older PHENIX measurement~\cite{ppg044}. However, the old data lack statistical significance at high transverse momenta.

Another way to look at the centrality dependence is the ratio of the yield in more central collisions to the yield in the most peripheral collisions, $R_{CP}$. The advantage is that common systematic uncertainties cancel, especially  the uncertainty of the energy scale of the detector. The centrality evolution of $R_{CP}$ is shown in Figure~\ref{fig:rcp}. Here, due to the slight suppression in central collisions and the clear enhancement in peripheral collisions, a strong centrality dependence is observed, with $R_{CP}^{0-20\%/60-88\%}\approx 0.6$ at high $p_T$.

\begin{figure}[htbp]
\begin{minipage}[hbt]{79mm}
\begin{center}
\includegraphics[width=\textwidth]{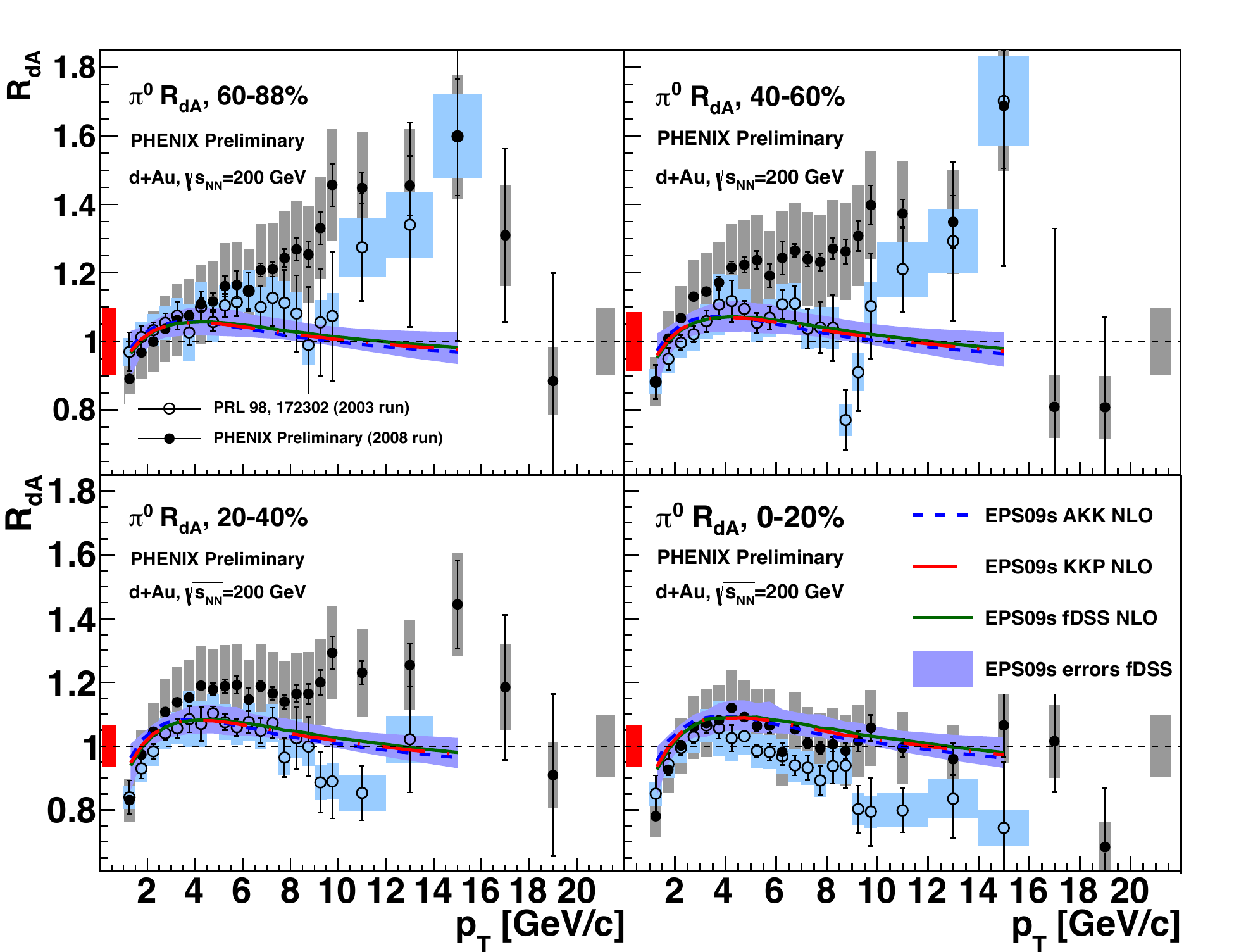}
\end{center}
\caption{Comparison of the measured $\pi^0$ $R_{dA}$ (dots) with EPS09s. The 2003 data~\cite{ppg044} are shown as well (circles). The boxes show the same systematic uncertainties as in Figure~\ref{fig:rda}. While EPS09s agrees with the data in the most central events, it does not describe them in the more peripheral ones.}
\label{fig:eps09scomp}
\end{minipage}
\hspace{\fill}
\begin{minipage}[hbt]{54mm}
\begin{center}
\includegraphics[width=0.94\textwidth]{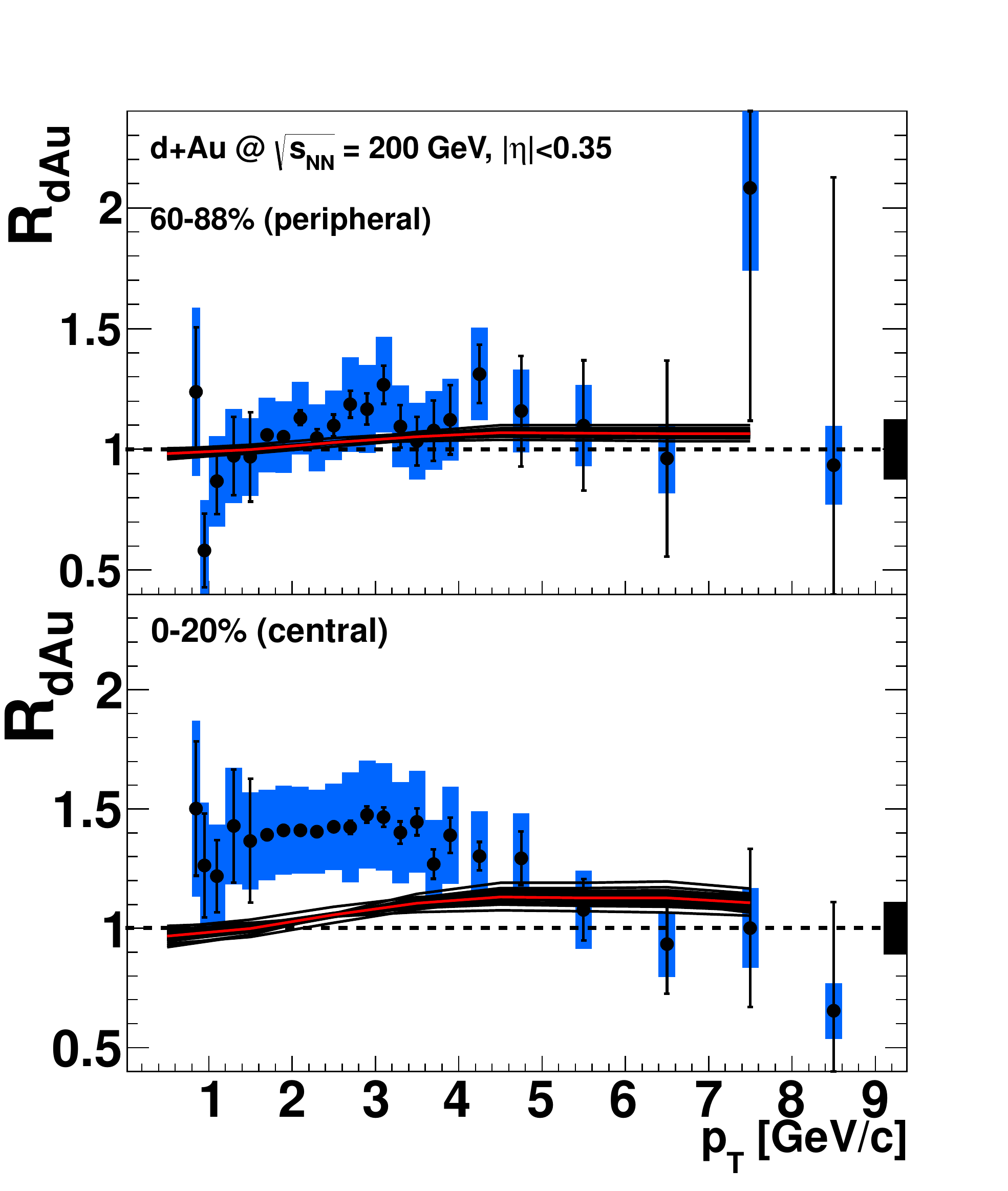}
\end{center}
\caption{Comparison of the measured heavy-flavor electron $R_{dA}$ (dots) with a centrality dependent EPS09 calculation, for peripheral (top) and central (bottom) collisions. The boxes show the same systematic uncertainties as in Figure~\ref{fig:rda}.}
\label{fig:singleel}
\end{minipage}
\end{figure}

A comparison of the $\pi^0$ data with the EPS09s nPDFs is shown in Figure~\ref{fig:eps09scomp}, both the 2003 and the new 2008 data are plotted to illustrate the agreement. It shows a good agreement in central $d$+Au collisions, while in peripheral collisions, the data and the EPS09s prediction are different. It is not clear, however, if the centrality definition and possible biases due to the aforementioned auto correlation are treated the same way in the data and in EPS09s.

\section{Electrons from Heavy-Flavor Decays}

\begin{figure}[htbp]
\end{figure}

Recently, PHENIX has measured centrality-dependent single electron spectra and $R_{dA}$ from heavy-flavor decays~\cite{ppg131}. For this measurement, all electrons from other sources such as $\pi^0$ Dalitz decay, decays of other light mesons that do not carry charm or beauty, and quarkonia are subtracted from the overall electron sample. A striking observation was an enhancement in the region of about 1\,GeV/$c$ $<$ $p_T$ $<$ 5\,GeV/$c$. In Figure~\ref{fig:singleel}, the single electron $R_{dA}$ is compared with a calculation using EPS09 for the nPDF modification, where the centrality dependence is included through the integrated longitudinal density of the Au nucleus (see~\cite{jpsipaper}). The calculation is based on PYTHIA~\cite{pythia} input spectra. Here, in peripheral collisions, the prediction from EPS09 agrees with the data while it is in disagreement in the enhancement region in central collisions. However, the calculation does not include $k_T$ broadening from parton scattering in the nucleus which would be expected to lead to an enhancement in this $p_T$ region.

\section{Summary}

PHENIX has measured the centrality dependence of the spectra and the nuclear modification factor in $d$+Au collisions at $\sqrt{s_{NN}}=200$\,GeV, for $\pi^0$, $\eta$, reconstructed jets, and single electrons from heavy-flavor decays. The jet measurement shows the same trend in centrality dependence as the measurement of $\pi^0$ and $\eta$, which is expected if the high $p_T$ hadrons are leading fragments from jets. With the current centrality definition, a strong centrality dependence is observed, where $R_{dA}$ is slightly suppressed ($R_{dA}\approx 0.85$) in central collisions and significantly enhanced ($R_{dA}\approx 1.3 - 1.4$) in peripheral collisions. This behavior is not explained by centrality dependent nuclear PDFs such as EPS09s. Further studies are necessary to understand if this new phenomenon comprises new physics effects in the nucleus or indicates potential bias in events which include detected jets or very high pT particles.
Electrons from heavy-flavor decays are modified in central $d$+Au collisions, for 1\,GeV/$c$ $<$ $p_T$ $<$ 5\,GeV/$c$, this enhancement disappears in peripheral collisions. This observation is in the region where Cronin enhancement might be present also for heavy quarks.


\section*{References}

\begin{thebibliography}{00} 
\bibitem{PHENIXwhitepaper}K.\,Adcox, et.al.,  Nucl.Phys. {\bf A757}  (2005) 184. 
\bibitem{Cronin}J.\,Cronin, et.al., Phys.Rev. {\bf D11} (1975) 3105.
\bibitem{eps09s}I.\,Helenius, K.\,Eskola, H.\,Honkanen, C.\,Salgado, JHEP {\bf 1207} (2012) 073.
\bibitem{ppg044}S.\,S.\,Adler, et.al., arXiv:nucl-ex/0610036.
\bibitem{gaussian} Y.-S.\,Lai, B.\,Cole, arXiv:0806.1499.
\bibitem{ppg131}A.\,Adare, et.al.,  arXiv:1208.1293.
\bibitem{jpsipaper}A.\,Adare, et.al., Phys.Rev.Lett.{\bf107} (2011) 142301.
\bibitem{pythia}T.\,Sj\"ostrand, et.al., Comput. Phys. Commun. {\bf 135} (2001) 238.

\end{thebibliography}
\end{document}